\documentclass[11pt]{article}
\usepackage[utf8]{inputenc}
\usepackage{listings}
\usepackage{color}
\usepackage{graphicx}
\usepackage{longtable}
\usepackage{float}
\usepackage{amsmath}
\usepackage{authblk}

\title{}
\date{\today}

\begin{document}

\title{Interplay of receptor memory and ligand rebinding}
\author{Thorsten Pr\"ustel} 
\author{Martin Meier-Schellersheim} 
\affil{Laboratory of Systems Biology\\National Institute of Allergy and Infectious Diseases\\National Institutes of Health}
\maketitle
\let\oldthefootnote\thefootnote 
\renewcommand{\thefootnote}{\fnsymbol{footnote}} 
\footnotetext[1]{Email: prustelt@niaid.nih.gov, mms@niaid.nih.gov} 
\let\thefootnote\oldthefootnote 
\abstract{
Rapid rebinding of molecular interaction partners that are in close proximity after dissociation leads to a dissociation and association kinetics that can profoundly differ from predictions based on bulk reaction models. The cause of this effect can be traced back to the non-Markovian character of the ligand's rebinding time probability density function, reflecting the fact that, for a certain time span, the ligand still 'remembers' the receptor it was bound to previously. In this manuscript, we explore the consequences of the hypothesis that initial binding and consecutive rebinding give rise to a bond lifetime density that is non-Markovian as well.  We study the combined effect of the two non-Markovian waiting time probability densities and show that even for very short times the decay of the fraction of occupied receptors deviates from an exponential. For long times, dissociation is slower than an exponential and the fate of the the steady-state bound receptor fraction critically depends on the extent of the deviation, relative to the rebinding time density, from the Markovian limit: The population of occupied receptors may either decay completely or assume a non-vanishing value for small and strong deviations, respectively. Furthermore, we point out the important role played by fractional calculus and demonstrate that the short- and long-time dynamics of the occupied receptors can be naturally expressed as well as easily obtained in terms of fractional differential equations involving the Riemann-Liouville derivative.
Our analysis shows that cells may exploit receptor memory as mechanism to dynamically widen the range of potential response-patterns to a given signal.
}
\section{Introduction}
Growing evidence \cite{Taeuber:2005, Taeuber_2:2005, Takahashi:2010p139, Ten_Wolde:2012} suggests that a ligand's diffusive motion results in a receptor-ligand interaction pattern and a network response further downstream that can deviate substantially from the picture based on deterministic first-order kinetics. A major cause of these differences has been identified as the non-Markovian rebinding time probability density function (pdf) that arises in stochastic, spatially-resolved accounts of the ligand's motion \cite{Takahashi:2010p139, Ten_Wolde:2012}. This rebinding time pdf may be considered as a 'deformed' version of the memoryless exponential waiting time pdf that corresponds to the first order kinetics of a deterministic mean-field description.
In fact, the rebinding time pdf $\psi_{\text{reb}}(t)$ for 1-dimensional diffusive motion in the presence of a partially absorbing boundary takes in the Laplace domain the form (see appendix A)
\begin{equation}\label{Rebinding-Laplace}
\mathcal{L}[\psi_{\text{reb}}(t)](s):=\tilde{\psi}_{\text{reb}}(s) := \int^{\infty}_{0}\psi_{\text{reb}}(t)e^{-st}dt = \frac{\kappa_{D}^{1/2}}{\kappa_{D}^{1/2} + s^{1/2}},
\end{equation}    
where $\kappa_{D} = \kappa^{2}_{a} /D$ is a parameter whose strength is determined by the ratio of the intrinsic association constant $\kappa_{a}$ to the diffusion constant $D$. Note that $\kappa_{D}$ has the dimension of an inverse time. 

By contrast, the exponential waiting time pdf that is commonly employed to describe the dissociation reaction (and association as well in a non-spatial deterministic approach),
\begin{equation}\label{Exponential_Waiting_Time_PDF}
\psi(t) = \kappa_{d} e^{-\kappa_{d} t},
\end{equation}
reads in the Laplace domain as 
\begin{equation}\label{Exponential-Laplace}
\tilde{\psi}(s) = \frac{\kappa_{d}}{\kappa_{d}+s},
\end{equation}
where $\kappa_{d}$ refers to the intrinsic dissociation constant.
Both density functions (Eqs.~\eqref{Rebinding-Laplace}, \eqref{Exponential-Laplace}) are representatives of a family of pdf that are parametrized by a parameter $0 < \sigma \leq 1$ \cite{AgmonWeiss1989, Mainardi:2008, TPMMS_2013JCP}
\begin{equation}\label{laplace_non_markov}
\tilde{\psi}(s) = \frac{\kappa^{\sigma}}{\kappa^{\sigma}+s^{\sigma}}.
\end{equation}
Any PDF with $\sigma < 1$ describes memory effects: Because a ligand, that just dissociated from a receptor, will remain close to that receptor for a certain time span it is more likely to bind the same receptor again. The spatial correlations between receptor and dissociated ligand translate to a time evolution of the ligand's (survival) probability that is non-local in time and hence give rise to a dynamics that can be naturally cast as a fractional differential equation. Fractional calculus \cite{Oldham, Miller:1993} plays a prominent role in anomalous diffusion and non-equilibrium statistical mechanics of complex systems \cite{Metzler:2000, Tachiya:2003, Metzler:2004, Sokolov:2005, Klafter, West:2014}, but it also naturally arises in the context of normal diffusion. To see this in more detail, let us recall that the rebinding PDF  $\psi_{\text{reb}}(t)$ of a single ligand is related to the survival probability of a molecule that undergoes a 1-dimensional diffusive motion in the presence of a partially absorbing boundary condition (BC) at $z_{0}=0$ by (cf. Appendix A) 
\begin{equation}\label{Waiting-Survival-Relation}
 \psi_{\text{reb}}(t) = -\frac{\partial S_{\text{rad}}(t\vert z_{0}=0)}{\partial t},
\end{equation}
Henceforth, we will use the notation $S(t):= S_{\text{rad}}(t\vert z_{0}=0)$ for convenience. From Eqs.~\eqref{Rebinding-Laplace}, \eqref{Waiting-Survival-Relation} and \eqref{Laplace-Survival-Probability} we can conclude that
\begin{equation}
1 - s\tilde{S}(s) = \kappa^{1/2}_{D}s^{1/2} \tilde{S}(s). 
\end{equation}
This relation reads in the time domain as
\begin{equation}\label{Time-Evolution-S}
\frac{\partial S(t)}{\partial t} = -\kappa^{1/2}_{D}D^{1-1/2}_{t} S(t),
\end{equation}
where $D^{1-\sigma}_{t}$ denotes the Riemann-Liouville (R-L) fractional differential operator that is defined as \cite{Oldham, Miller:1993, Metzler:2000}
\begin{equation}\label{R-L-Derivative}
_{0}D^{1-\sigma}_{t}f(t) := D^{1-\sigma}_{t}f(t) := \frac{1}{\Gamma(\sigma)}\frac{\partial}{\partial t}\int^{t}_{0} \frac{f(\tau)}{(t-\tau)^{1-\sigma}}d\tau, \quad 0 < \sigma < 1.
\end{equation}
The definition of the R-L derivative (Eq.~\eqref{R-L-Derivative}) makes it evident that the time evolution of $S(t)$ (Eq.~\eqref{Time-Evolution-S}) is indeed governed by an equation that is non-local in time. Eq.~\eqref{Time-Evolution-S} generalizes the well-known Markovian time evolution
\begin{equation}
\frac{\partial S(t)}{\partial t} = -\kappa_{D}S(t)
\end{equation}
that leads to a Poissonian form of the rebinding pdf. Solutions of evolution equations that take the form of a fractional relaxation law \cite{Nonnenmacher:1995, Metzler:2000, metzler:2002}
\begin{equation}\label{Fractional-Relaxation}
\frac{\partial \Phi(t)}{\partial t} = -\kappa^{\sigma}D^{1-\sigma}_{t} \Phi(t)
\end{equation}
are given in terms of Mittag-Leffler (M-L) functions \cite{Erdelyi:Bateman, Mainardi:2000} $\Phi(t) = E_{\sigma}[-(\kappa t)^{\sigma}]$. M-L functions may be thought of a generalized exponentials that are defined by the series expansion 
\begin{equation}\label{M-L-F-Def}
E_{\alpha}(-x) = \sum^{\infty}_{n=0}\frac{(-x)^{n}}{\Gamma(n\alpha + 1)}, 
\end{equation}
which makes it evident that for $\alpha=1$ one recovers an exponential $E_{1}(-t) = \exp(-t)$.
The Laplace transform is 
\begin{equation}\label{Laplace-M-L}
\mathcal{L}\bigg[E_{\sigma}(-(\kappa t)^{\sigma}) \bigg] = \frac{1}{s}\frac{s^{\sigma}}{s^{\sigma}+\kappa^{\sigma}}
\end{equation}
and hence it follows that the pdf Eq.~\eqref{laplace_non_markov} is the Laplace transform of the negative time derivative of the Mittag-Leffler (M-L) function $E_{\sigma}[-(\kappa t)^{\sigma}]$ \cite{Mainardi:2008}:
\begin{equation}
\psi(t) = -\frac{\partial E_{\sigma}[-(\kappa t)^{\sigma}]}{\partial t}.
\end{equation} 
Because $E_{1/2}[-t^{1/2}] = \exp(t)\text{erfc}(\sqrt{t})$ we immediately obtain from Eq.~\eqref{Time-Evolution-S} the known expression for $S(t)$, cf.~Appendix A, Eq.~\eqref{Classical-Surv-Prob}.
For small arguments, the M-L function behaves like a Kohlrausch-Williams-Watts (KWW) function, i.e. a stretched exponential $E_{\alpha}[-t^{\alpha}]\sim \exp(-t^{\alpha})$ \cite{Metzler:2000, metzler:2002} (cf. Eq.~\eqref{M-L-F-Def}), while for large arguments it decays like an inverse power law. More precisely, one has \cite{Mainardi:2000, Metzler:2000}
\begin{eqnarray}
E_{\alpha}[-t^{\alpha}] &=& \frac{t^{-\alpha}}{\Gamma(1 - \alpha)} - \frac{t^{-2\alpha}}{\Gamma(1-2\alpha)} +\ldots \quad  t\rightarrow \infty.\label{M-L-Long-Time}
\end{eqnarray}

We have seen that fractional evolution equations (Eq.~\eqref{Time-Evolution-S}) that reflect deviations from a Markovian waiting time pdf naturally appear in the decription of ligand rebinding, even in the case of normal diffusion.
However, the reverse process, dissociation is usually taken into account by a memoryless pdf for the bond lifetime, even in particle-based stochastic approaches (however, note, for instance, Refs.~\cite{Nonnenmacher:1995, Haenggi:2004}). Here, we suggest to abandon the assumption of the Markovian property. A deviation from a memoryless bond lifetime pdf is motivated by several lines of experimental evidence suggesting that the receptor state does not remain unaltered upon initial rebinding, for instance due to receptor deformation or conformational changes, the formation of signaling intermediates, involvement of the actin skeleton and cluster formation and other mechanisms \cite{Nonnenmacher:1995, Haenggi:2004, Zar:2007}. In the picture that we will consider, the bound state undergoes a diffusion-like 'motion' in an abstract internal 'space' whose properties are dynamically defined by the local biochemistry near the receptor. We will show that the interplay of these two different diffusive processes alters substantially the apparent dissociation kinetics and can even lead to behavior that has no counterpart in models that only take into account either the non-Markovian character of the ligand rebinding or the bond lifetime alone.

Let us recall how an altered receptor state may lead to a bond lifetime pdf that is sufficiently well described by Eq.~\eqref{laplace_non_markov} \cite{AgmonWeiss1989, Nonnenmacher:1995}. To this end, one assumes
that the receptor modifications imply a bond lifetime pdf that is given by a superposition of exponentials
\begin{equation}
\psi(t) = \sum_{n}\phi_{n}e^{-\kappa_{n}t}.
\end{equation}
One further assume that the sum can be approximated by an integral
\begin{equation}\label{Def_Psi_Integral}
\psi(t) = \int^{\infty}_{0}\phi(\kappa)e^{-\kappa t}d\kappa.
\end{equation}
Then, using mild assumptions, one can derive (cf.~Appendix B) that, asymptotically, the pdf behaves like a power-law
\begin{equation}\label{Psi_Power_Law}
\psi(t) \propto \frac{\Gamma(1 + \sigma)}{t^{1 + \sigma}}.
\end{equation}
Employing Tauberian theorems \cite{Feller-2, Klafter} one can show that
\begin{equation}\label{asymp_psi}
\tilde{\psi}(s) \sim 1 - (\tau s)^{\sigma},
\end{equation}
where $\tau$ denotes a constant having dimension of time.
Now, for large times, i.e. small $s$ or for small $\tau$ we finally may rewrite Eq.~(\ref{asymp_psi}) as Eq.~(\ref{laplace_non_markov}), identifying $\tau^{-1}=\kappa$.

Therefore, we will henceforth assume that the bond lifetime pdf is indeed given by Eq.~\eqref{laplace_non_markov}. We consider the following system: Ligands diffuse in a 3 dimensional space that is semi-infinite, bounded by a two-dimensional plane homogeneously covered with receptors. We adopt the self-consistent stochastic mean-field theory
approach of Ref.~\cite{Taeuber:2005} that allows to track the motion of the ligand accurately, while the receptors are described in a non-spatial fashion by the fraction of occupied
receptors. We point out that the formalism is flexible enough to include also a spatially-resolved description of the receptors \cite{Taeuber_2:2005}. Here, however, we aim at emphasizing the 
consequences of a non-Markovian bond lifetime density and its interplay with the rebinding pdf and therefore limit our analyses to a non-spatial receptor description.  
\section{Theory}
First, we focus on the Markovian case \cite{Taeuber:2005}.
We consider a membrane that extends infinitely in the $x-y$ plane ($z=0$) and that is uniformly covered with receptors. We assume that all ligands are initially bound. The fraction of occupied receptors is denoted by $p(t)=\mathcal{R}(t)/\mathcal{R}_{0}$ where $\mathcal{R}(t)$ and $\mathcal{R}_{0}$ refer to the density of bound receptors at time $t$ and the constant mean receptor surface density per unit area, respectively.
Its time evolution is governed by \cite{Taeuber:2005}
\begin{equation}\label{eq_of_motion}
\frac{dp(t)}{dt} = -k_{\text{off}}p(t) + k_{\text{on}}\rho(0, t)[1 - p(t)],
\end{equation}
where $\rho(z,t)$ describes the concentration of ligands at a distant $z$ from the membrane. 
To solve Eq.~\eqref{eq_of_motion} one has to express the ligand density $\rho(0, t)$ and hence the rebinding rate 
\begin{equation}
\gamma(t) = k_{\text{on}}\rho(0, t)[1-p(t)]
\end{equation}
in terms of $p(t)$. To this end, we follow the self-consistent mean-field theory approach described in Refs.~\cite{Taeuber:2005} and assume that $p(0) \ll 1$ and hence $p(t) \ll 1$ for all times $t$ so that the rebinding rate reduces to $\gamma(t) = k_{\text{on}}\rho(0, t)$. We now aim at expressing $\gamma(t)$ in an alternative form that incorporates spatially-resolved information about the diffusive motion of the ligand. The fraction of occupied receptors which is converted to the unbound state between time $\tau$ and $\tau + d\tau$ is given by $k_{\text{off}}p(\tau)d\tau$. A dissociated ligand begins to undergo an effectively one-dimensional diffusive motion that starts at the mebrane $z=0$ between time $\tau$ and $\tau + d\tau$. Let $\mathcal{S}(t-\tau\vert z =0)$ denote the probability that a ligand is still unbound when the time span $t-\tau$ has elapsed since the dissociation. Then, the rate of absorption is given by
by the rebinding time pdf (see Appendix C)
\begin{equation}\label{Rebinding-Time-Survival-Many}
\psi_{\text{reb}}(t) := R_{\text{rad}}(t-\tau\vert z_{0} = 0) := -\frac{\mathcal{S}(t-\tau\vert z_{0} = 0)}{\partial t}
\end{equation}
and it follows that the rebinding rate can be written as convolution
\begin{equation}\label{rebinding_rate}
\gamma(t) = k_{\text{off}}\int^{t}_{0}d\tau p(\tau)R_{\text{rad}}(t-\tau\vert 0).  
\end{equation}
Note that, despite the formal resemblance of Eqs.~\eqref{Waiting-Survival-Relation} and \eqref{Rebinding-Time-Survival-Many}, we emphasize that the context of both relations is quite different, because here we consider a many-molecule system, instead of a single ligand-receptor pair. Also, the self-consistent mean-field theory approach is more flexible and can be applied to more general cases \cite{Taeuber_2:2005}.

As it stands, Eq.~\eqref{eq_of_motion} describes memoryless dissociation based on the exponential waiting time pdf Eq.~\eqref{Exponential_Waiting_Time_PDF}. To introduce a non-Markovian intrinsic bond lifetime, we perform the substitution 
\begin{equation}\label{Substitution-Rule}
k_{\text{off}}\rightarrow s\frac{\tilde{\psi}(s)}{1-\tilde{\psi}(s)}
\end{equation}
in the Laplace domain, where $\tilde{\psi}(s)$ is given by Eq.~\eqref{laplace_non_markov} with $\kappa = k_{\text{off}}$. Clearly, for $\sigma = 1$, one recovers $s\tilde{\psi}(s)/[1-\tilde{\psi}(s)] = k_{\text{off}}$. 
The substitution rule Eq.~\eqref{Substitution-Rule} has been utilized before \cite{AgmonWeiss1989, TPMMS_2013JCP}. 
However, Refs.~\cite{AgmonWeiss1989, TPMMS_2013JCP} have not identified the action of the operation of Eq.~\eqref{Substitution-Rule} in the time domain. Using \cite{Metzler:2000}
\begin{equation}\label{Laplace-R-L}
\mathcal{L}[D^{1-\sigma}_{t}f(t)](s) = s^{1-\sigma}\tilde{f}(s),
\end{equation}
it turns out that it corresponds to
\begin{equation}\label{substitution_rule_time}
-k_{\text{off}}p(t) \longrightarrow -k^{\sigma}_{\text{off}} D^{1-\sigma}_{t}p(t)
\end{equation}
where $D^{1-\sigma}_{t}p(t)$ refers to the R-L derivative (Eq.~\eqref{R-L-Derivative}).
Now, combining Eqs.~\eqref{Laplace-M-L}, \eqref{eq_of_motion}, \eqref{rebinding_rate}, \eqref{Laplace-R-L}, \eqref{substitution_rule_time} and \eqref{Laplace-Expression-R-Rad} we obtain
\begin{equation}\label{Fractional-Receptor-Laplace}
\tilde{p}(s) = s^{-1}p(0) - k^{\sigma}_{\text{off}}s^{-\sigma} \frac{s^{1/2}}{s^{1/2}+\kappa^{1/2}_{D}}\tilde{p}(s),
\end{equation}
or, equivalently,
\begin{equation}\label{Fractional-Receptor-Time}
p(t) = p(0) - k^{\sigma}_{\text{off}}D^{1-\sigma}_{t}\bigg\lbrace E_{1/2}[-(\kappa_{D}t)^{1/2}] \ast p(t) \bigg\rbrace,
\end{equation}
where $\ast$ denotes a convolution integral. 

Eq.~\eqref{Fractional-Receptor-Time} represents a natural generalization of a rate equation approach, abandoning the assumptions of both a memoryless rebinding and dissociation waiting time pdf. In fact, considering the limit $\sigma\rightarrow 1$ and substituting $ E_{1/2}[-(\kappa_{D}t)^{1/2}] \rightarrow E_{1}[-(\kappa_{D}t)^{1}] = \exp(-\kappa_{D}t)$, we obtain from
Eq.~\eqref{Fractional-Receptor-Time} the deterministic Markovian mean-field result
\begin{equation}
p(t) = p(0)\bigg[e^{-(\kappa_{D}+k_{\text{off}})t}\frac{k_{\text{off}}}{\kappa_{D}+k_{\text{off}}} + \frac{\kappa_{D}}{\kappa_{D}+k_{\text{off}}}\bigg].
\end{equation}
Next, we turn to the general case and focus on the short and long time limit.
We will show that in most cases the dynamics is described by the same fractional relaxation law (Eq.~\eqref{Fractional-Relaxation}), only the order of the R-L derivative changes from case to case. Hence, as a benefit, the solutions can be virtualy read off, a fact which underlines the power and elegance of the chosen formalism.
\section{Results}  
\subsection{Solution for short times $t\rightarrow 0$}
For very short times, the effect of the rebinding should be negligible. This expectation is well reflected by Eq.~\eqref{Fractional-Receptor-Time}: Because short times imply that $E_{1/2}[-(\kappa_{D}t)^{1/2}] \rightarrow 1$, it follows that the equation of motion assumes the form of a fractional relaxation law,
\begin{equation}\label{Fractional-Short-Time-Relaxation}
\frac{dp(t)}{dt} = -k^{\sigma}_{\text{off}} D^{1-\sigma}_{t}p(t),
\end{equation}    
for which we can swiftly read off the solution (cf. Eqs.~\eqref{Fractional-Relaxation} and \eqref{M-L-F-Def})
\begin{eqnarray}
p(t) &=& p(0)E_{\sigma}[-(k_{\text{off}}t)^{\sigma}] \label{Small_Times_R_Occupancy-I}\\
&=& p(0)\bigg[1 - \frac{(k_{\text{off}}t)^{\sigma}}{\Gamma(1+\sigma)} + \frac{(k_{\text{off}}t)^{2\sigma}}{\Gamma(1+2\sigma)} - \ldots\bigg] \quad \text{for}\, t\rightarrow 0. \label{Small_Times_R_Occupancy-II}
\end{eqnarray}
Note that to arrive at Eq.~\eqref{Fractional-Short-Time-Relaxation}, we invoked the commutation relation $[D^{1-\sigma}_{t}, \partial /\partial t]f(t) = 0$ that does hold \textit{if} $f(0)=0$. Since in the case considered here we have $f(t) = \int^{t}_{0}p(\tau)d\tau$, the condition is fullfilled. As we know from the behavior of the M-L function (and can see from Eq.~\eqref{Small_Times_R_Occupancy-II}), for short times the fraction of occupied receptors decays like a stretched exponential $p(t)\sim p(0)\exp[-(k_{\text{off}}t)^{\sigma}]$. 

We emphasize that the solution given by Eq.~\eqref{Small_Times_R_Occupancy-I} is not only valid for short times, but in addition in parameter regimes in which $\kappa^{1/2}_{D} = \mathcal{R}_{0}k_{\text{on}}/\sqrt{D}$ becomes small, i.e.~when either the receptor density is small or when rebinding effects are weak, i.e. the ratio of on-rate to diffusion constant becomes small. Because for $\sigma = 1$ the corresponding M-L function in Eq.~\eqref{Small_Times_R_Occupancy-I} is a decaying exponential $E_{1}[-(k_{\text{off}}t)^{1}] = \exp(-k_{\text{off}}t)$, we recover
the result from \cite{Taeuber:2005} for short times. However, here we see that abandoning the assumption of a memoryless bond lifetime leads to an decay that deviates from an exponential even for short times or for small $\kappa_{D}$.
\subsection{Solutions for long times $t\rightarrow\infty$}
Now we will show that the combined effect of rebinding and receptor memory gives rise to behavior that has no counterpart in cases in which one uses only one of the two non-Markovian densities. The long time limit, or equivalently $s\rightarrow 0$, implies that (cf. Eq.~\eqref{Laplace-M-L})
\begin{equation}
\mathcal{L}\bigg\lbrace E_{1/2}[-(\kappa_{D}t)^{1/2}] \bigg\rbrace \rightarrow \frac{1}{\kappa^{1/2}_{D} s^{1/2}}.
\end{equation}
As a result, Eq.~\eqref{Fractional-Receptor-Time} again takes the form of a fractional relaxation law, as in the case of short times, but the order of the R-L derivative has changed (cf. with Eq.~\eqref{Fractional-Short-Time-Relaxation})
\begin{equation}\label{Fractional-Long-Time-Relaxation}
\frac{dp(t)}{dt} = -\frac{k^{\sigma}_{\text{off}}}{\kappa^{1/2}_{D}} D^{1-\sigma+1/2}_{t}p(t).
\end{equation}
Here, we again emphasize that the found relaxation law is not only valid for long times, but also for regimes in which $\kappa_{D}$ becomes large, i.e.~where rebinding effects becomes strong.
Furthermore, the form of Eq.~\eqref{Fractional-Long-Time-Relaxation} makes it evident that there is now a crucial difference whether $\sigma < 1/2$ or $\sigma > 1/2$ i.e.~whether the bond lifetime is to a larger extent non-Markovian than the rebinding time pdf or not.
In this sense, diffusion sets the scale for what should be considered a strong or weak deviation from memoryless behavior. Note that this scale is not absolute: If the ligand undergoes anomalous diffusion \cite{Metzler:2000, Metzler:2004, Tachiya:2003, Sokolov:2005}, the scale will not be $1/2$ anymore.

We now consider several special cases separately. First, let us focus on the Markovian limit $\sigma = 1$. Then, we immediately obtain from Eq.~\eqref{Fractional-Long-Time-Relaxation}
\begin{equation}
p(t) = E_{1/2}\bigg[-\bigg(\frac{k^{2}_{\text{off}}}{\kappa_{D}}t\bigg)^{1/2}\bigg] = \exp\bigg[\bigg(\frac{k^{2}_{\text{off}}}{\kappa_{D}}t\bigg)\bigg]\text{erfc}\bigg[\sqrt{\frac{k^{2}_{\text{off}}}{\kappa_{D}}t}\bigg],
\end{equation}
which is the result given in Ref.~\cite{Taeuber:2005, Taeuber_2:2005}, as expected. In particular, we observe a power law decay for long times $p(t\rightarrow \infty)\sim 1/\sqrt{\pi t}$.

Next, we analyze the case $1 > \sigma > 1/2$. It follows that $\sigma - 1/2 > 0$ and hence we can read off again immediately the solution from Eq.~\eqref{Fractional-Long-Time-Relaxation}
\begin{eqnarray}\label{Long_Times_Sigma_Larger}
p(t) &=& p(0) E_{\sigma - 1/2}\bigg[-k^{\sigma}_{\text{off}}/\kappa^{1/2}_{D}t^{\sigma-1/2}\bigg] = \nonumber \\ 
&& p(0)\bigg[\frac{\kappa^{1/2}_{D}}{ k^{\sigma}_{\text{off}}\Gamma[1-(\sigma - 1/2)] t^{\sigma - 1/2}} - \frac{\kappa_{D}}{k^{2\sigma}_{\text{off}}\Gamma[1-2(\sigma - 1/2)] t^{2\sigma - 1}} +\ldots\bigg], \qquad\quad
\end{eqnarray}
where we employed the expansion of the M-L function for large arguments Eq.~\eqref{M-L-Long-Time}.
We note that the fraction of occupied receptors will completely decay for long times, as in the Markovian case $\sigma = 1$. However, it can also be seen from Eq.~\eqref{Long_Times_Sigma_Larger} that for general $1/2 < \sigma < 1$ the decay can become arbitrarily slow for $\sigma \rightarrow 1/2$ and hence $\sigma = 1/2$ seems to play a special role that we are going to consider now.

As previously, the asymptotic solution $p(t\rightarrow\infty)$ for the limiting case $\sigma = 1/2$ can easily be obtained from the general relaxation law Eq.~\eqref{Fractional-Long-Time-Relaxation} and using $E_{0}(-x) = 1/(1+x)$
\begin{equation}
p(t\rightarrow\infty)=p(0)\frac{\kappa^{1/2}_{D}}{\kappa^{1/2}_{D} + k^{1/2}_{\text{off}}}.
\end{equation}
Thus, we find that the steady-state receptor occupancy is \textit{not} vanishing in the long time limit, in stark contrast to the case of an exponential decay of the intrinsic bond lifetime.
We point out that Agmon and Weiss found a corresponding behavior for the case of a single pair in one dimension \cite{AgmonWeiss1989}. But as mentioned before, the formalism adopted here is more general \cite{Taeuber:2005, Taeuber_2:2005}. Moreover, we establish a connection to 3D/2D membrane binding as well as fractional calculus and the crucial role of the fractional relaxation law.   

Finally, we turn to the case $\sigma < 1/2$. This is the only case for which we cannot read off the solution immediately. The reason is that the nature of the R-L derivative changes qualitatively.
In fact, let be $\alpha = 1/2-\sigma$, then $\alpha > 0$ and one has \cite{Metzler:2000}
\begin{equation}
D^{1+\alpha}_{t}p(t) = \frac{1}{\Gamma(1-\alpha)}\frac{\partial^{2}}{\partial t^{2}}\int^{t}_{0}\frac{p(\tau)}{(t-\tau)^{\alpha}}d\tau.
\end{equation}
Therefore, we Laplace transform Eq.~\eqref{Fractional-Long-Time-Relaxation} (or, alternatively, start from \eqref{Fractional-Receptor-Laplace}) and obtain 
\begin{eqnarray}
\tilde{p}(s) &=& \kappa^{1/2}_{D}p(0)\frac{s^{\sigma-1}}{\kappa^{1/2}_{D}s^{\sigma} + k^{\sigma}_{\text{off}}s^{1/2}} \nonumber\\
&=& \frac{p(0)}{s} \bigg[1 - \frac{1}{1 + \kappa^{1/2}_{D}/k^{\sigma}_{\text{off}}s^{\sigma-1/2}}\bigg].
\end{eqnarray}
The corresponding expression in the time domain follows from Eq.~\eqref{Laplace-M-L}
\begin{equation}
p(t) = p(0)\bigg\lbrace1 - E_{1/2 - \sigma}\bigg[ -\kappa^{1/2}_{D}/k^{\sigma}_{\text{off}}t^{1/2 - \sigma}\bigg]\bigg\rbrace
\end{equation}
In particular, we find that 
\begin{equation}
\lim_{t\rightarrow\infty}p(t) = p(0)
\end{equation}
the initial fraction of occupied receptors is restored completely.
\section*{Appendix A} 
We consider a molecule undergoing 1-dimensional diffusive motion in the presence of a partially absorbing BC. The pdf $g_{\text{rad}}(z, t\vert z_{0})$ that yields the probability $g_{\text{rad}}(z, t\vert z_{0})dz$ to find the molecule between $z$ and $z+dz$ at time $t$, provided that it was initially located at $z_{0}$ at time $t_{0} = 0$, is the GF of the diffusion equation
\begin{equation}\label{Diffusion-Equation}
\frac{\partial g_{\text{rad}}(z, t\vert z_{0})}{\partial t} = D \frac{\partial^{2} g_{\text{rad}}(z, t\vert z_{0})}{\partial z^{2}},
\end{equation}
subject to the radiation BC \cite{collins1949diffusion, Sano_Tachiya:1979}
\begin{equation}\label{Radiation-BC}
D\frac{\partial g_{\text{rad}}(z, t\vert z_{0})}{\partial z}\vert_{z=0} = \kappa_{a}g_{\text{rad}}(z = 0, t\vert z_{0}).
\end{equation}
The GF takes in the Laplace domain the form \cite[p. 358]{carslaw1986conduction}
\begin{equation}\label{GF-Laplace}
\tilde{g}_{\text{rad}}(z, s\vert z_{0}) = \frac{e^{-\sqrt{s/D}\vert z - z_{0}\vert}}{2\sqrt{Ds}}  + \frac{e^{-\sqrt{s/D}(z + z_{0})}}{2\sqrt{Ds}} - \frac{\kappa_{a}e^{-\sqrt{s/D}( z - z_{0})}}{D\sqrt{s}[\sqrt{s} + \kappa_{a}/\sqrt{D}]}.    
\end{equation}
The survival probability $S_{\text{rad}}(t\vert z_{0})$ is defined as \cite{Sano_Tachiya:1979}
\begin{equation}
S_{\text{rad}}(t\vert z_{0}) = \int^{\infty}_{0} g_{\text{rad}}(z, t\vert z_{0}) dz,
\end{equation}
and hence one finds
\begin{equation}\label{Laplace-Survival-Probability}
\tilde{S}_{\text{rad}}(s\vert z_{0} = 0) = \frac{1}{s}\frac{s^{1/2}}{s^{1/2} + \kappa^{1/2}_{D}},
\end{equation}
where $\kappa_{D} = \kappa^{2}_{a}/D$.
Eq.~\eqref{Laplace-Survival-Probability} becomes in the time domain \cite[Eq.~(29.3.43)]{abramowitz1964handbook}
\begin{equation}\label{Classical-Surv-Prob}
S_{\text{rad}}(t\vert z_{0} = 0) = \exp(\kappa_{D}t)\text{erfc}(\sqrt{\kappa_{D}t}).
\end{equation}
Furthermore, from Eqs.~\eqref{Diffusion-Equation} and \eqref{Radiation-BC} one may derive the relation
\begin{equation}\label{S-and-Rad-BC}
-\frac{S_{\text{rad}}(t\vert z_{0})}{\partial t} = \kappa_{a}g_{\text{rad}}(z = 0, t\vert z_{0}).
\end{equation}
Finally, using Eqs.~\eqref{GF-Laplace} and \eqref{S-and-Rad-BC} 
we obtain 
\begin{equation}
\tilde{\psi}_{\text{reb}}(s) = \mathcal{L}\bigg[-\frac{\partial S_{\text{rad}}(t\vert z_{0}=0)}{\partial t}\bigg] = \frac{\kappa^{1/2}_{D}}{\kappa^{1/2}_{D} + s^{1/2}}
\end{equation}
\section*{Appendix B}
In this Appendix we recall \cite{AgmonWeiss1989, Nonnenmacher:1995, Klafter} how the form of the density $\phi(\kappa)$, that appears in Eq.~\eqref{Def_Psi_Integral}, can be derived from rather mild assumptions. 
First, one assumes that a typical rate $\kappa$ is characterized by an Arrhenius factor
\begin{equation}\label{Arrhenius}
\kappa \propto e^{-\frac{E}{kT}},
\end{equation}
where $k, T$ refer to the Boltzmann constant and the absolute temperature, respectively.
The activation energies $E$ themselves are assumed to be distributed according to an Arrhenius factor too
\begin{equation}\label{energy_dist}
\phi(E) \propto e^{-\frac{E}{E_{0}}},
\end{equation}
where $E_{0}$ denotes a energy reference scale.
Using Eq.~\eqref{Arrhenius}, one may express $E$ by $\kappa$ and substitute the obtained expression in Eq.~\eqref{energy_dist} to find
\begin{equation}\label{form_kappa_dist}
\phi(\kappa) \propto \kappa^{\sigma},
\end{equation}
where $\sigma := kT/E_{0}$. 
Eq.~\eqref{Def_Psi_Integral} defines a Laplace transform where $\kappa$ plays the role of the time variable. Hence, knowing Eq.~\eqref{form_kappa_dist}, one may employ $\mathcal{L}[t^{\alpha-1}]=\Gamma(\alpha)/s^{\alpha},\,\alpha > 0$ \cite[Eq.~(29.3.7)]{abramowitz1964handbook} to finally arrive at Eq.~\eqref{Psi_Power_Law}.
\section*{Appendix C}
We consider first a single ligand-receptor pair.
The rate of recombination $R_{\text{rad}}(t\vert z_{0}) = -\partial S_{\text{rad}}(t\vert z_{0})/\partial t$ can be written as convolution \cite{Pedersen:1980}
\begin{equation}\label{Recombination-Convolution}
R_{\text{rad}}(t\vert z_{0}) = \Theta R_{\text{abs}}(t\vert z_{0}) + (1-\Theta)\int^{t}_{0}R_{\text{abs}}(\tau\vert z_{0}) R_{\text{rad}}(t-\tau\vert \Delta)d\tau \quad\text{for}\,\Delta\rightarrow 0
\end{equation}
where $R_{\text{abs}}(t\vert z_{0})$ refers to the rate of first encounters (first-passage time pdf)
\begin{equation}
R_{\text{abs}}(t\vert z_{0}) = -\frac{S_{\text{abs}}(t\vert z_{0})}{\partial t},
\end{equation}
and $S_{\text{abs}}(t\vert z_{0})$ denotes the survival probability in the presence of an absorbing BC.
Furthermore, $\Delta$ is a microscopic length scale and $\Theta$ denotes the probability that a single molecule pair reaction takes place when they encounter.
However, here we are interested in a situation where we consider many receptors. Eq.~\eqref{Recombination-Convolution} remains valid \cite{Taeuber:2005, Taeuber_2:2005}, but the probability $\Theta$ has to be adapted properly. Following Refs.~\cite{Taeuber:2005, Taeuber_2:2005}, we define
\begin{equation}\label{Definition-Theta}
\Theta = \frac{\mathcal{R}_{0}k_{\text{on}}\Delta}{D}.
\end{equation}
Next, we Laplace transform Eq.~\eqref{Recombination-Convolution} to arrive at
\begin{equation}\label{Recombination-Rate-Laplace}
\tilde{R}_{\text{rad}}(s\vert z_{0}) = \Theta \tilde{R}_{\text{abs}}(s\vert z_{0}) + (1-\Theta)\tilde{R}_{\text{abs}}(s\vert z_{0}) \tilde{R}_{\text{rad}}(s\vert \Delta).
\end{equation}
Setting $z_{0} = \Delta$, we obtain from Eq.~\eqref{Recombination-Rate-Laplace} an expression for $\tilde{R}_{\text{rad}}(s\vert z_{0}=\Delta)$ and substite it back in Eq.~\eqref{Recombination-Rate-Laplace} to find
\begin{equation}
\tilde{R}_{\text{rad}}(s\vert z_{0}) = \frac{\Theta \tilde{R}_{\text{abs}}(s\vert z_{0})}{1 - (1-\Theta)\tilde{R}_{\text{abs}}(s\vert \Delta)}.
\end{equation}
Using $\tilde{R}_{\text{abs}}(s\vert z_{0}) = e^{-\frac{z_{0}}{\sqrt{D}}\sqrt{s}}$ \cite{Taeuber:2005} and Taylor-expanding up to first order in $\Delta$ yields
\begin{equation}
\tilde{R}_{\text{rad}}(s\vert z_{0}) = \frac{\Theta \tilde{R}_{\text{abs}}(s\vert z_{0})}{\Theta + (1-\Theta)\Delta/\sqrt{D}s^{1/2}}
\end{equation}
Finally, we take the limit $\Delta \rightarrow 0$ (cf. Eq.~\eqref{Definition-Theta})
\begin{equation}
\lim_{\Delta \rightarrow 0} \frac{\Delta}{\Theta} = \frac{D}{k_{\text{on}}\mathcal{R}_{0}}
\end{equation}
to arrive at 
\begin{equation}\label{Laplace-Expression-R-Rad}
\tilde{\psi}_{\text{reb}}(s) = \tilde{R}_{\text{rad}}(s\vert z_{0}=0) = \frac{1}{1 + \kappa^{-1/2}_{D} s^{1/2}}, \quad \kappa^{1/2}_{D} := \mathcal{R}_{0}k_{\text{on}}/\sqrt{D}.
\end{equation} 
In particular, this result means that the rate of recombination is the negative time derivative of the M-L function $E_{1/2}[-\mathcal{R}_{0}k_{\text{on}}/\sqrt{D}t^{1/2} ]$.
\section*{Acknowledgments}
This research was supported by the Intramural Research Program of the NIH, National Institute of Allergy and Infectious Diseases. 

\end{document}